\documentclass[aps,twocolumn,showpacs, superscriptaddress, groupedaddress,nofootinbib]{revtex4}  
\usepackage{graphicx}  
\usepackage{dcolumn}   
\usepackage{bm}        
\usepackage{amssymb, amsmath}
\usepackage{wasysym}
\usepackage[usenames,dvipsnames,svgnames]{xcolor}
\definecolor{darkblue}{rgb}{0,0.1,0.5}
\definecolor{darkgreen}{rgb}{0,0.5,0.2}
\definecolor{seablue}{rgb}{0,0.2,0.6}
\usepackage{hyperref}
\hypersetup{
    colorlinks=true,       
    linkcolor=darkblue,          
    citecolor=BrickRed,        
    filecolor=magenta,      
    urlcolor=MidnightBlue           
}
\usepackage[all]{hypcap} 
\usepackage{lipsum, color}

\def\beq{\begin{equation}}
\def\eeq{\end{equation}}
\begin{document}
\widetext
\leftline{DESY 17-148}
\leftline{YITP-SB-17-41}


\title{
New LHC bound on low-mass diphoton resonances
}
\author{Alberto Mariotti}
\affiliation{Theoretische Natuurkunde and IIHE/ELEM, Vrije Universiteit Brussel, and International Solvay
Institutes, Pleinlaan 2, B-1050 Brussels, Belgium}
\author{Diego Redigolo}
\affiliation{Raymond and Beverly Sackler School of Physics and Astronomy, Tel-Aviv University, Tel-Aviv
69978, Israel}
\affiliation{Department of Particle Physics and Astrophysics, Weizmann Institute of Science, Rehovot 7610001,Israel}
\author{Filippo Sala}
\affiliation{DESY, Notkestra{\ss}e 85, D-22607 Hamburg, Germany}
\affiliation{LPTHE, UMR 7589 CNRS, 4 Place Jussieu, F-75252, Paris, France}
\author{Kohsaku Tobioka}
\affiliation{Raymond and Beverly Sackler School of Physics and Astronomy, Tel-Aviv University, Tel-Aviv
69978, Israel}
\affiliation{Department of Particle Physics and Astrophysics, Weizmann Institute of Science, Rehovot 7610001,Israel}
\affiliation{C.N. Yang Institute for Theoretical Physics, Stony Brook University, Stony Brook, NY 11794-3800}

\begin{abstract}
\noindent We derive a new bound on diphoton resonances using inclusive diphoton cross section measurements at the LHC, in the so-far poorly constrained mass range between the $\Upsilon$ and the SM Higgs. This bound sets the current best limit on axion-like particles that couple to gluons and photons, for masses between 10 and 65 GeV. 
We also estimate indicative sensitivities of a dedicated diphoton LHC search in the same mass region, at 7, 8 and 14 TeV
.
As a byproduct of our analysis, we comment on the axion-like particle interpretation of the CMS excesses in low-mass dijet and diphoton searches.

\end{abstract}

\maketitle

\section{Introduction}

Searches for two body decays of heavy resonances led to fundamental discoveries in the history of particle physics such as the J/$\psi$ \cite{PhysRevLett.33.1404,PhysRevLett.33.1406}, the $\Upsilon$ \cite{PhysRevLett.39.252} and the $Z$ boson \cite{Arnison:1983mk}. An extensive program is currently looking for higher mass resonances at the LHC in various final states (see \cite{Craig:2016rqv} for a complete list).

Despite the high background rates, advances in data-driven background estimates guarantee good sensitivities to discover/exclude such peak signals. A marvellous proof of the high performance of resonance searches at the LHC is the recent discovery of the Standard Model (SM) Higgs boson in the diphoton channel~\cite{Aad:2012tfa,Chatrchyan:2012xdj}.  

As a matter of fact, the current LHC search program is mostly tailored to probe new resonances of mass higher than roughly 100 GeV.
This is the result of a general theoretical bias towards heavy new physics (NP) and of the common belief that either previous collider experiments (UA1, UA2, LEP and Tevatron) and/or Higgs coupling fits (through the decay of the Higgs into two new particles) put constraints on lighter resonances that are stronger than the LHC capabilities.
On the experimental side, going to low 
masses poses the challenge of 
 looking for resonances with a mass below the sum of the cuts on the transverse momentum $(p_T)$ of the decay products. 

The aim of this letter is to go beyond these common beliefs and to motivate the LHC collaborations to look for resonances down to the smallest possible mass.
We first derive a new bound (of 10 - 100 pb) on the diphoton signal strength of a new resonance in the mass range between the $\Upsilon$ and the SM Higgs. This new bound comes from inclusive diphoton cross section measurements at ATLAS~\cite{Aad:2012tba,Aaboud:2017vol} and CMS~\cite{Chatrchyan:2014fsa}. Assuming zero knowledge about the background, we simply impose that the NP events are less than the total measured events plus twice their uncertainty.

We show how this conservative procedure sets already the strongest existing constraint on axion-like particles (ALPs) with mass between 10 and 65 GeV. We finally estimate the indicative reaches on the diphoton 
signal strengths that could be attainable
by proper searches at the LHC, up to its high luminosity (HL) phase, and interpret their impact on the ALP parameter space.

\section{Axion-like particles in diphotons}
When a $U(1)$ global symmetry (which can be the subgroup of some larger global symmetry $\mathcal{G}$) is spontaneously broken in the vacuum, then a massless Nambu-Goldstone boson (NGB) arises in the low energy spectrum. If the $U(1)$ symmetry is only approximate, the NGB gets a mass $m_a$ and it becomes a pseudo-Nambu-Goldstone boson (pNGB), often called axion-like particle (ALP). The mass $m_a$ of the pNGB is a technically natural parameter which depends on the explicit breaking of the $U(1)$ global symmetry, and is smaller than the associated NP scale $M_{\rm NP} \sim 4\pi f_a$, where $f_a$ is the scale of spontaneous breaking. In particular $m_a$ can be smaller than the SM Higgs mass without any fine-tuning price. 

The axial couplings of the pNGB to SM gauge bosons can be written as
\begin{equation}
\mathcal{L}_{\text{int}}= \frac{a}{4\pi f_a}\left[ \alpha_s c_3 G\tilde{G}+ \alpha_2 c_2 W\tilde{W}+ \alpha_1 c_1 B\tilde{B}\right],
\label{eq:La123}
\end{equation}
where $\alpha_1=5/3 \alpha'$ is the GUT normalized $U(1)_Y$ coupling constant, $a$ is the canonically normalized pNGB field, and the coefficients $c_i$ encode the Adler-Bell-Jackiw (ABJ) anomalies of the global $U(1)$ with $SU(3)$ and $SU(2)\times U(1)_Y$. Further couplings of the pNGB with the SM Higgs and/or with the SM fermions can be set to zero if these fields are not charged (or very weakly charged) under the global $U(1)$. 

As one can see from Eq.~\eqref{eq:La123}, the strength of the couplings of the pNGB is controlled by its decay constant $f_a$. As we will show, the phenomenology of the pNGB becomes of interest for this study, and more in general for present colliders, for $f_a \sim0.1-10$~TeV. Decay constants in this range are ubiquitous in popular theoretical frameworks addressing the naturalness of the EW scale, like low-scale Supersymmetry (SUSY) and Compositeness.\footnote{
String theory constructions could provide an extra motivation for  ALPs. However, the expected values of $f_a$ in string models like~\cite{Svrcek:2006yi,Arvanitaki:2009fg,Cicoli:2012sz} are order of magnitudes too high for being phenomenologically interesting at colliders. 
Similarly, solutions of the strong CP problem based on a QCD axion~\cite{Peccei:1977hh,Peccei:1977ur,Weinberg:1977ma,Wilczek:1977pj} with a decay constant $f_a$ at the TeV scale are hard to conceive (see however~\cite{Fukuda:2015ana,Dimopoulos:2016lvn}).
}
Note that generically we expect that other fields associated to the $U(1)$ spontaneous breaking (e.g. the radial mode) should have a mass $\lesssim 4 \pi f_a$. Hence in the lower extreme of the range for $f_a$ other signatures associated to the BSM theory could be accessible at the LHC.

Supersymmetry (SUSY) and its breaking predict on general grounds the existence of an $R$-axion~\cite{Nelson:1993nf}, pNGB of the $U(1)_R$ symmetry, potentially accessible at the LHC if the SUSY scale is sufficiently low~\cite{Bellazzini:2017neg}.
In this context the couplings to gauge bosons of Eq.~(\ref{eq:La123}) are realized naturally from ABJ anomalies between $U(1)_R$ and the SM gauge group, while the couplings to SM fermions and Higgses can be set to zero with a well-defined $R$-charge assignment ($R_H = 0$ in the notation of~\cite{Bellazzini:2017neg}). In composite Higgs models, attempts of fermionic UV completions point to the need of non-minimal cosets (see e.g. \cite{Barnard:2013zea,Ferretti:2013kya,Ferretti:2014qta}), which in turn imply the existence of pNGBs lighter than the new confinement scale. 
See~\cite{Ferretti:2016upr} for recent work about these pNGBs, and~\cite{Belyaev:2015hgo} for a systematic classification of the cosets structures that give rise to pNGBs that couple to both gluons and EW gauge bosons.

A common feature of both SUSY and Composite Higgs models is that the QCD anomaly receives an irreducible contribution from loops of colored states, like gluinos and/or tops, which are generically chiral under the spontaneously broken $U(1)$. As a consequence one typically expects $c_3\neq0$, unless model dependent cancellations occur. 
In conclusion, $f_a \sim0.1-10$~TeV and  $c_3\neq 0$ in a broad class of SUSY and Composite Higgs models, so that $a$ is copiously produced in $pp$ collisions at the LHC. For this reason we believe that our study applies to a wide range of theoretically motivated ALP models.

From a phenomenological point of view, ALPs of interest for this study have received much attention as mediators of simplified Dark Matter models (see for example the recent~\cite{Banerjee:2017wxi}). 
Finally, ALPs can exist if Strong Dynamics is present at some scale \cite{Kilic:2009mi}. In such a case, having $f_a \sim0.1-10$~TeV would be a phenomenological assumption not motivated by any naturalness consideration. 


For $m_a \lesssim m_h$, the relevant two body decays of $a$ are in diphotons and dijets, with widths
\begin{equation}
\Gamma_{gg}=K_g\,\frac{\alpha_s^2c_3^2}{8 \pi^3}  \frac{m_a^3}{f_a^2},
\qquad\qquad \Gamma_{\gamma\gamma}=\frac{\alpha_{\text{em}}^2c_\gamma^2}{64 \pi^3}  \frac{m_a^3}{f_a^2},
\label{eq:widths}
\end{equation}
where $c_\gamma = c_2 + 5 c_1/3$, and where both $\alpha_s$ and $\alpha_{\text{em}}$ are computed at the mass of $m_a$. We encode the higher-order QCD corrections in $K_g = 2.1$~\cite{Djouadi:2005gj}.
Unless $c_{1,2} \gtrsim 10^2 c_3$, the width into gluons is the dominant one. The total width $\Gamma_{\text{tot}}$ is typically very narrow, for example for $f_a\gtrsim 100\text{ GeV}$ and $c_i \sim O(1)$ one obtains $\Gamma_\text{tot}/m_a\lesssim 10^{-3}$.

For simplicity, we do not study the phenomenology associated to the $Z\gamma$ decay channel, which is anyhow open only for $m_a > m_Z$, and phenomenologically more relevant than $\gamma\gamma$ only for specific values of $c_1$ and $c_2$.

\section{Current Searches}

A new resonance decaying in two jets or two photons is probed at colliders by looking at the related invariant mass distributions, possibly in addition with extra objects, either SM or BSM (see e.g.~\cite{Delgado:2016arn,Ellwanger:2017skc}) depending on the production mechanism.
We summarise and discuss here the most relevant searches for light resonances at the LHC, and refer to Appendix~\ref{app:previoussearches} for a more complete list and a discussion of the existing searches and of diphoton cross section measurements, at the LHC, Tevatron, LEP and Sp$\bar{\rm p}$S.

\begin{itemize}
\item[$\diamond$] Dijet resonances down to 50 GeV have been recently looked for by CMS~\cite{Sirunyan:2017nvi}.  In order to overcome the trigger on the jet $p_T$'s, CMS has a strong cut on the total hadronic activity $H_T$.  Recoiling against the hard jet, the resonance is boosted and its decay products collimated. For this reason advanced jet substructure techniques were essential to reconstruct the dijet resonance inside a single ``fat'' jet \cite{Dolen:2016kst,Moult:2016cvt}.

The CMS low-mass dijet limits are given on the inclusive dijet signal strength of a $q\bar{q}$-initiated resonance $\sigma_{q\bar{q}}^{\rm CMS}$.
We recast them for a gluon initiated resonance as
\begin{equation}
\sigma_{gg}^{\rm our} = \sigma_{q\bar{q}}^{\rm CMS}\cdot \frac{ \epsilon_{H_T}^{q\bar{q}}}{\epsilon_{H_T}^{gg}}\,,
\label{eq:ggfromqq}
\end{equation}
where $\epsilon_{H_T}^{q\bar{q}}$ and $\epsilon_{H_T}^{gg}$ are the efficiencies of the cut in hadronic activity $H_T>650\text{ GeV}$.\footnote{
We thank Phil Harris for private communications on~\cite{Sirunyan:2017nvi}.
}
These are estimated from simulations\footnote{
Throughout this paper we use FeynRules 2.0 \cite{Alloul:2013bka}, MadGraph 5 v2 LO \cite{Alwall:2011uj,Alwall:2014hca} with the default pdf set, Pythia 8.1 \cite{Sjostrand:2006za,Sjostrand:2007gs}, DELPHES 3 \cite{deFavereau:2013fsa} and MadAnalysis 5 \cite{Conte:2012fm}. The MLM matching \cite{Alwall:2007fs} is performed to include matrix element correction to ISRs. 
} of a $gg$ and a $q\bar{q}$ initiated scalar signals (including matching up to 2 jets and detector simulation).
We take the efficiency ratio in Eq.~\eqref{eq:ggfromqq} to be constant and equal to $0.08$, which is the value that we find at $m_a = 80$ GeV. Accounting for the $m_a$ dependence introduces variations up to $20\%$ within the mass range $50-125$ GeV.
The fact that the efficiency ratio is roughly constant in $m_a$  can be understood observing that $\sqrt{\hat{s}}$ is always dominated by the cut of $H_T >$ 650 GeV, which is much larger than any of the values of $m_a$ of our interest.

\item[$\diamond$] Existing diphoton searches are inclusive and extend to a lower invariant mass of $65$ GeV~\cite{Aad:2014ioa,CMS-PAS-HIG-14-037,Khachatryan:2015qba, CMS-PAS-HIG-17-013}, where the two photons satisfy standard isolation and identification requirements.  

The ATLAS diphoton search at 8 TeV \cite{Aad:2014ioa} is the one extending down to $65$ GeV. The bound is given in term of the diphoton ``fiducial'' cross-section $\sigma^{\text{fid}}=\sigma^{th}\cdot \epsilon_S/C_X$.  $C_X$ is a model independent number that we take from \cite{Aad:2014ioa} and encodes the detector acceptance once the kinematical cuts are already imposed ($C_X\simeq 0.6$ in the mass range of our interest).\footnote{We thank Liron Barak for private communications on~\cite{Aad:2014ioa}.} 
To extract the efficiency $\epsilon_S$ we simulated the signal for the ALP model in Eq.~\eqref{eq:La123} accounting for all the cuts of \cite{Aad:2014ioa}. 

The CMS searches at 8 and 13 TeV \cite{CMS-PAS-HIG-14-037,CMS-PAS-HIG-17-013} provide the bound on the theoretical signal strength for a resonance with the same couplings of the SM Higgs but lighter mass. Since gluon fusion is the dominant production mechanism for a SM Higgs in the low mass range~\cite{Heinemeyer:2013tqa}, we take the CMS result as a bound on the theoretical diphoton signal strength of our ALP. 

\end{itemize}

\section{New Bound and LHC sensitivities\quad
from $\gamma\gamma$ cross-section measurements}

\begin{table*}[t!]
\centering
\begin{tabular}{|c|cccccccccccc|}
\hline
$m_a$ in GeV&10 &20 &30 &40 & 50&60&70&80&90&100&110&120\\
\hline
$\epsilon_S$ for $\sigma_{ 7 \text{TeV}}$ ATLAS \cite{Aad:2012tba}&0&0.008&0.022&0.040&0.137&0.293&0.409&0.465&0.486&0.533&0.619&0.637\\
$\epsilon_S$ for $\sigma_{ 7 \text{TeV}}$ CMS \cite{Chatrchyan:2014fsa}&0&$0.002$&$0.010$&$0.020$&$0.030$&$0.058$&$0.156$&$0.319$&$0.424$&$0.499$&$0.532$&$0.570$\\
$\epsilon_S$ for $\sigma_{ 8 \text{TeV}}$ ATLAS \cite{Aaboud:2017vol}&0&0.0007&0.008&0.014&0.024&0.037&0.071&0.233&0.347&0.419&0.452&0.484\\
$\epsilon_S$ for $\sigma_{ 2 \text{TeV}}$ CDF \cite{Aaltonen:2011vk,PhysRevLett.110.101801}&0.001& 0.007& 0.026& 0.143& 0.212& 0.241& 0.276& 0.275& 0.283& 0.3&
0.319&0.327\\
$\epsilon_S$ for $\sigma_{ 2 \text{TeV}}$ D0 \cite{Abazov:2010ah}&0& 0.002& 0.008& 0.018& 0.114& 0.169& 0.208& 0.21& 0.217& 0.234&
0.244& 0.252\\
\hline
\end{tabular}
\caption
{Signal efficiencies for the 7 TeV and 8 TeV cross-section measurements at the LHC \cite{Aad:2012tba,Aaboud:2017vol,Chatrchyan:2014fsa} and at the Tevatron \cite{Aad:2012tba,Aaboud:2017vol} for a resonance produced in gluon fusion. \label{tab:efficiencies}}
\end{table*}

Here we extract a new bound on diphoton resonances from inclusive diphoton measurements at the LHC and at Tevatron, and we present the projected LHC sensitivities.

\paragraph{New bound from measurements.}
The papers~\cite{PhysRevLett.110.101801,Aad:2012tba,Aaboud:2017vol,Chatrchyan:2014fsa} provide tables of the measured differential diphoton cross sections per invariant mass bin, $\text{d}\sigma_{\gamma\gamma}/\text{d} m_{\gamma\gamma}$, together with their relative statistical ($\Delta_{\rm stat}$) and systematical ($\Delta_{\rm sys}$) uncertainties.
We derive a conservative bound on the theoretical signal strength $\sigma^{\text{th}}_{\gamma\gamma}$ of a diphoton resonance by imposing 
\begin{equation} 
\sigma^{\text{th}}_{\gamma\gamma} (m_a) \lesssim \left[m_{\gamma\gamma}^{\rm Bin}\cdot \frac{\text{d}\sigma_{\gamma\gamma}}{\text{d} m_{\gamma\gamma}}\left(1+2\Delta_{\rm tot}\right)\right] \cdot \frac{1}{\epsilon_S(m_a)}\ ,
\label{eq:pigbound}
\end{equation}
where $\Delta_{\rm tot} = \sqrt{\Delta_{\rm sys}^2+\Delta_{\rm stat}^2}$, $m_{\gamma\gamma}^{\text{Bin}}$ is the size of the bin containing $m_a$, and $\epsilon_S$ is the signal efficiency accounting for the kinematical and the isolation cuts of the photons.

At a given center of mass energy $s$, we derive $\epsilon_S$ as
\begin{equation}
\epsilon_S(m_a) = \frac{\sigma^{\text{MCcuts}}_{\gamma\gamma} (m_a,s)}{C_s\,\sigma^{\text{LO}}_{\gamma\gamma} (m_a,s)}\,.
\label{eq:efficiency}
\end{equation}
$\sigma_{\gamma\gamma}^{\text{LO}}(m_a, s)$ is the LO gluon fusion cross section, derived using the gluon pdf from~\cite{Martin:2009iq}, multiplied by the LO branching ratio into $\gamma\gamma$ computed from Eq.~(\ref{eq:La123}).
We also compute a total ``simulated'' diphoton signal strength $\sigma^{\text{MCtot}}_{\gamma\gamma}$, which includes matching up to 2 jets, by a Monte Carlo (MC) simulation of the signal for the ALP model in Eq.~(\ref{eq:La123}). We find that $\sigma^{\text{LO}}_{\gamma\gamma}$ reproduces up to a constant factor $C_s$ the shape of $\sigma^{\text{MCtot}}_{\gamma\gamma}$ for $m_{\gamma \gamma}\gtrsim 60\text{ GeV}$ (i.e. sufficiently far from the sum of the minimal detector $p_T$ cuts on the photons). A constant factor  $C_s\equiv\sigma^{\text{MCtot}}_{\gamma\gamma}(s)/\sigma^{\text{LO}}_{\gamma\gamma}(s)$ is hence included in Eq.~\eqref{eq:efficiency} and we obtain $C_{7\,\text{TeV}}\simeq C_{8\,\text{TeV}}\simeq 0.85$ while $C_{2\,\text{TeV}}\simeq1$ at the Tevatron center of mass energy.
The signal strength after cuts $\sigma^{\text{MCcuts}}_{\gamma\gamma}$ is obtained by the MC simulations imposing
on the events samples the relevant cuts for each of the experimental search.

To validate our procedure with a measured quantity, we simulate the SM diphoton background and verify that it reproduces well the experimental diphoton cross section measurements of~\cite{Aad:2012tba,Aaboud:2017vol}.
We refer the reader to Appendix~\ref{app:binning} for more details on our derivation of $\epsilon_S(m_a)$, including validations. We list in Table~\ref{tab:efficiencies} the efficiencies as a function of $m_a$ for the various cross section measurements.

We stress that, for very light mass values, a NP resonance can pass the cuts on the photon $p_T$'s by recoiling against a jet, which is not vetoed since the cross section measurements are inclusive. This is reflected in the efficiencies of the signal which are non vanishing (thought small) also in the region of very low resonance mass.

\begin{figure}[t!]
\includegraphics[width=0.49\textwidth]{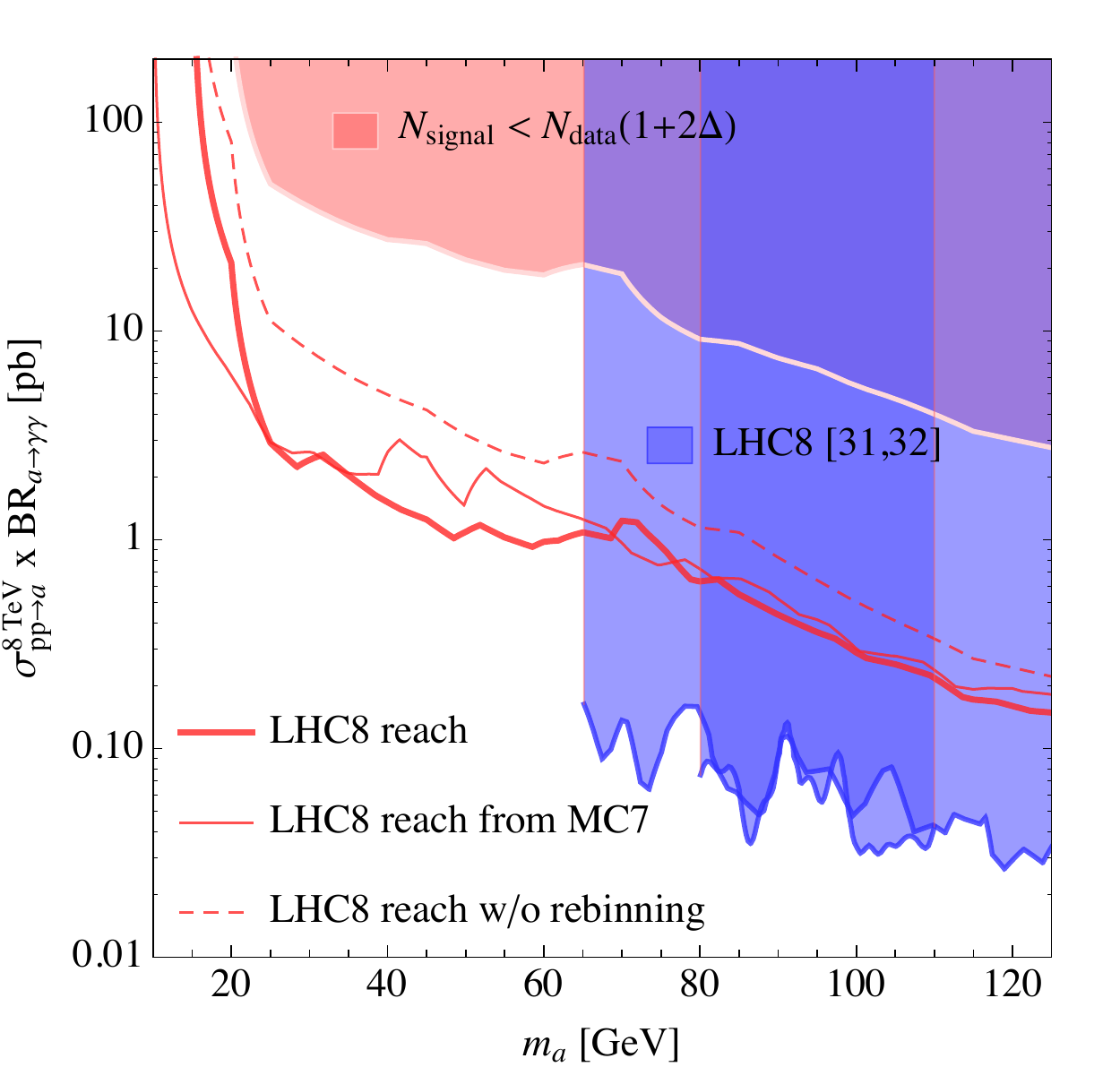}
\caption{
Bounds (shaded) and expected sensitivities (lines) on the diphoton signal strength of a resonance produced in gluon fusion, at 8 TeV. More details in the text. 
\label{fig:modelind}
}
\end{figure}

Our final results are shown in Fig.~\ref{fig:modelind}, where the conservative bound extracted from 8 TeV ATLAS data \cite{Aaboud:2017vol} using Eq.~\eqref{eq:pigbound} is compared against the existing 8 TeV searches at ATLAS~\cite{Aad:2014ioa} and CMS~\cite{CMS-PAS-HIG-14-037}. 


\paragraph{Sensitivities from measurements.}
An expected sensitivity $\sigma^{\text{sens}}_{\gamma\gamma}$ can be derived by assuming the measured cross section to be dominated by the SM diphoton background, and requiring the signal to be within the $2\Delta_{\rm tot}$ variation of the background:
\begin{equation}
\sigma^{\text{sens}}_{\gamma\gamma}(m_a) = \left[m_{\gamma\gamma}^{\rm Bin}\cdot \frac{\text{d}\sigma_{\gamma\gamma}}{\text{d} m_{\gamma\gamma}}\cdot 2\Delta_{\rm tot}\right] \cdot \frac{1}{\epsilon_S (m_a)}.
\label{eq:sensitivity}
\end{equation}
The sensitivities we present in Fig.~\ref{fig:modelind} as thick continuous and dashed lines correspond to two different choices of $m_{\gamma\gamma}^{\rm Bin}$, and both correspond to 8 TeV data with integrated luminosity 20.2~fb$^{-1}$~\cite{Aaboud:2017vol}.

The most conservative sensitivity between the two corresponds to the binning given directly in the ATLAS 8 TeV cross section measurement \cite{Aaboud:2017vol}, where the mass bins have a size of 30 to 10 GeV in the region of our interest. 
A better sensitivity is obtained by reducing the bin size $m_{\gamma\gamma}^{\rm Bin}$ down to the invariant mass resolution obtained from the ATLAS and CMS ECAL energy resolution on a single photon, that we extract from \cite{2006NIMPA.568..601A} and \cite{deFavereau:2013fsa}, and which leads to mass bins of size $\simeq 3$~GeV for values of $m_a$ below the sum of the minimal $p_T$ cuts of the photons (see Appendix~\ref{app:binning} for more details).
Since the signal is narrow, the number of signal events in the bin is not affected. The number of background $N_{\rm bkg}$ events is instead reduced and the sensitivity increased assuming that the errors scale as $\sqrt{N_{\rm bkg}}$.\footnote{
The CMS sensitivities using different binning in Fig.~\ref{fig:modelind} are very close in the 75-100 GeV range.
This is because in this mass range CMS reports its measurement in 5 GeV bins, comparable to the ECAL mass resolution of $\sim$ 2.5 GeV, while in other mass ranges (and in the ATLAS measurements) the bin sizes vary between 10 and 40 GeV.
} This scaling holds for statystical errors and we assume the same scaling for systematical ones. The assumption is motivated by the scaling of some of the systematics (e.g. those associated to poor statistics in control regions) and by the fact that the CMS cross section measurements~\cite{Chatrchyan:2014fsa} do not separate statistical from systematical uncertainties.

\paragraph{Sensitivities adding MC input, up to 14 TeV.}
Now we discuss how to rescale the sensitivities from lower energies $\sqrt{s}_{\rm low}$ to higher energies $\sqrt{s}_{\rm high}$.
To rescale the diphoton background we first obtain, from MC simulations, $\sigma^{\rm MC}_{\rm low}$ and $\sigma^{\rm MC}_{\rm high}$.
These are the SM diphoton cross sections at $\sqrt{s}_{\rm low}$ and $\sqrt{s}_{\rm high}$ after the cuts of the cross
section measurements at $\sqrt{s}_{\rm low}$ are imposed.
We then take $ \sigma_{\gamma\gamma, \rm high}^{\rm bkg} = \sigma_{\gamma\gamma, \rm low}^{\rm bkg} \sigma^{\rm MC}_{\rm high}/\sigma^{\rm MC}_{\rm low} $, where $\sigma_{\gamma\gamma, \rm low}^{\rm bkg}$ is extracted from the experimental measurements.
The total relative uncertainties for the background are rescaled as the squared root of the total number of events so that $\Delta_{\rm high}= \sqrt{L_{\rm low}/L_{\rm high}}\sqrt{\sigma^{\rm MC}_{\rm low}/\sigma^{\rm MC}_{\rm high}} \,\Delta_{\rm low}$. Finally we also account for the different efficiencies for the signal going from $\sqrt{s}_{\rm low}$ to $\sqrt{s}_{\rm high}$. All in all, starting from Eq.~\eqref{eq:sensitivity} we get

\begin{equation}
\sigma^{\text{sens}}_{\gamma\gamma, \rm high}(m_a)
= \sqrt{\frac{L_{\rm low}}{L_{\rm high}} \cdot \frac{\sigma^{\rm MC}_{\rm high}}{\sigma^{\rm MC}_{\rm low}}}
\cdot \frac{\epsilon_S^{\rm low}}{\epsilon_S^{\rm high}}
\cdot \sigma^{\text{sens}}_{\gamma\gamma, \rm low}(m_a)\ .
\label{eq:sensitivityMC}
\end{equation}
We show it in Fig.~\ref{fig:modelind} for the extrapolation of the ATLAS reach from $\sqrt{s}_{\rm low} = 7$ TeV and 4.9~fb$^{-1}$ of data to $\sqrt{s}_{\rm high} = 8$ TeV and 20.2~fb$^{-1}$ of data (thus with the cuts of the ATLAS7 measurement~\cite{Aad:2012tba}). The overlap (in the region where the difference in the cuts matters less) between the 8 TeV sensitivities and the rescaled ones from 7 TeV is a nice consistency check of our procedure. We find an analogous agreement between the two 14 TeV sensitivities derived from 7 and 8 TeV data, as shown in Appendix~\ref{app:extramaterial}.

\section{Discussion}
Our sensitivities assume the uncertainties from MC modelling to be subdominant with respect to the ones associated to the measurement. However, this might not be the case in the entire mass range~(see e.g. \cite{Aad:2012tba, Aaboud:2017vol,Chatrchyan:2014fsa}) and a better control on the MC modelling might be necessary.
The current MC uncertainty can be read off e.g. \cite{Aaboud:2017vol}, and can be as large as 40\% for $m_{\gamma\gamma}$ below the minimal $p_T$ cuts of the photons
(see also \cite{Kamenik:2016tve} for a discussion of the challenges of background modelling in the context of high mass diphoton resonances).
While the relatively good agreement of the MC modelling with the observed data would in principle make a discovery possible for large enough signal cross sections, the large MC uncertainties are a limiting factor to the discovery potential of a resonance search below the minimal $p_T$ cuts for the photons.

On the theory side this motivates an improvement in the diphoton MC's, while on the analysis side it pushes to extend the data-driven estimates of the background to lower $m_{\gamma\gamma}$, reducing further the associated uncertainties and thus improving the limits. Data-driven estimates of the SM background were indeed used in the ATLAS 8 TeV analysis \cite{Aad:2014ioa}, and we believe their effectiveness is at the origin of the discrepancy between our 8 TeV sensitivities and the actual ATLAS limits. As shown in Fig.~\eqref{fig:modelind} the discrepancy amounts to a factor of $\sim 5$.\footnote{We checked further differences between Ref.~\cite{Aad:2014ioa} and the procedure used here, such as a finer categorisation of the diphoton final states as in~\cite{Aad:2012tfa}, and a fully unbinned analysis. We find that they can affect the sensitivity at most by 20 - 40\%. } 

The experimental challenge of going to lower invariant masses is ultimately related to lowering the minimal cuts  $p_{T 1,2}^{\rm min}$ on the two photon $p_T$'s and/or relax the photon isolation requirement $\Delta R \gtrsim 0.4$, where $\Delta R \equiv \sqrt{\Delta\phi^2 + \Delta\eta^2}$ is the photon separation. 
Indeed by simple kinematics we get the strict lower bound on $m_{\gamma\gamma}$
\begin{equation}
m_{\gamma\gamma}>\Delta R\cdot \sqrt{p_{T1}^{\rm min} p_{T2}^{\rm min}},\label{eq:Filbound}
\end{equation}
where we used $m_{\gamma\gamma}^2= 2 p_{T1} p_{T2} (\cosh\Delta\eta - \cos\Delta\phi)$ that for small $\Delta\phi$ and $\Delta\eta$ is $m_{\gamma\gamma}^2\simeq \Delta R^2\cdot p_{T1} p_{T2} $. 
This absolute lower bound on $m_{\gamma\gamma}$ explains why in Fig.~\ref{fig:modelind} the 8 TeV reach derived from ATLAS7, which has the lowest $p_{T1,2}^{\rm min}$, can reach lower $m_{\gamma\gamma}$ than the ones derived from ATLAS8 measurements.

From Eq.~\eqref{eq:Filbound} we conclude that in order to extend the diphoton resonant searches to lower invariant masses one would have to lower either $p_{T1,2}^{\rm min}$ or $\Delta R $. Both these possibilities deserve further experimental study.

A first possible strategy would be to require a hard ISR jet in the diphoton analysis, along the way of what was done in the recent CMS search for low-mass dijet resonances \cite{Sirunyan:2017nvi}. The hard jet requirement would raise the $p_T$ of the resonance recoiling against it, collimating the two photons and hence posing the challenge of going to smaller $\Delta R$. In this kinematical regime, the two photons would look like a single photon-jet~\cite{Ellis:2012sd,Ellis:2012zp}
and it would be interesting to study if substructure techniques similar to those used in \cite{Sirunyan:2017nvi} for a dijet resonances can be applied to such an object.

A second strategy would be to lower the photon $p_{T1,2}^{\rm min}$. This, however, poses well-known problems with the SM background, like the larger backgrounds from QCD processes (see e.g.~\cite{Jaeckel:2015jla}) and the challenge of recording, storing, and processing so many events.\footnote{We thank Antonio Boveia and Caterina Doglioni for many clarifications on these matters.
}
One might handle the high data-rate and long-term storage challenge with the data scouting/Trigger-object Level Analysis methods~\cite{
CMS-DP-2012-022,
Khachatryan:2016ecr,
ATLAS-CONF-2016-030,
Sirunyan:2016iap,
Aaboud:2016leb
} where, rather than storing the full detector data for a given event, one stores only a necessary subset. 
Alternatively, one could accomodate lower trigger thresholds by recording full events for only a fixed fraction of the data~\cite{Khachatryan:2016bia,Aaboud:2016leb}, with \textit{prescaled} triggers, and/or setting aside these data for processing and analysis later~\cite{CMS-DP-2012-022,Aad:2014aqa} (data parking/delayed stream). 
Such techniques have already been used in searches for dijet signals~\cite{Aad:2014aqa,Khachatryan:2016ecr,ATLAS-CONF-2016-030,Sirunyan:2016iap}, where one is similarly interested in localized deviations from smooth, data-driven background estimates.

The quantitative comparison of the reach of these different possibilities for low-mass diphoton resonances  goes beyond the scope of this paper, but we do encourage the ATLAS and CMS collaborations to take steps in these directions.

\section{Impact on ALP parameter space}

To determine the diphoton signal strength $\sigma^{\text{th}}_{\gamma\gamma}$ that enters the bound in Eq.~(\ref{eq:pigbound}) and that should be compared with the sensitivities in Eqs.~(\ref{eq:sensitivity}) and~(\ref{eq:sensitivityMC}), we multiply the tree level $pp$ cross section by a constant $K$-factor $K_\sigma = 3.7$~(see Appendix~\ref{app:signalXS} for more details) and we use the widths of Eq.~(\ref{eq:widths}).

\begin{figure}[t!]
\includegraphics[width=0.49\textwidth]{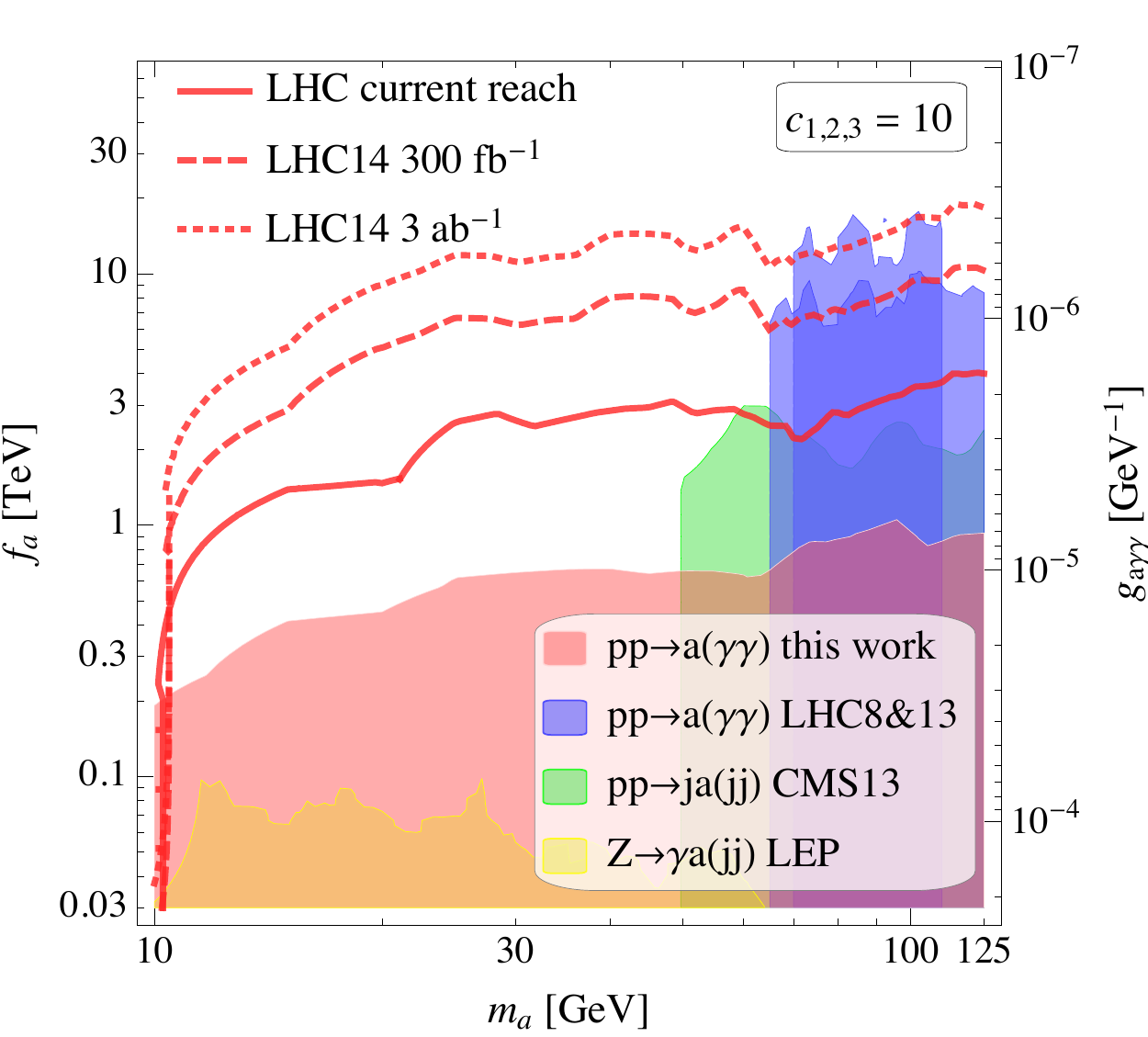}
\caption{Shaded: constraints on the ALP parameter space from existing collider searches at LEP~\cite{Adriani:1992zm} 
and the LHC~\cite{Aad:2014ioa,CMS-PAS-HIG-14-037,Sirunyan:2017nvi,CMS-PAS-HIG-17-013} (see text for our rescaling of the CMS dijet bound~\cite{Sirunyan:2017nvi}), and from the bound derived in this work using the data in~\cite{Aad:2012tba,Aaboud:2017vol,Chatrchyan:2014fsa}. Lines: our LHC sensitivities at 8 and 14~TeV. \label{fig:ALP}}
\end{figure}

In Fig.~\ref{fig:ALP} we show how the different searches at the LHC, at Tevatron and at LEP constrain the ALP decay constant $f_a$ for a given value of the ALP mass $m_a$. We fix for reference the anomalies to their GUT inspired value $c_1=c_2=c_3=10$. On the right $y$-axes, we write the pNGB coupling to photons in a notation inspired by the QCD axion,  as $g_{a\gamma\gamma}=\frac{\alpha_{\text{em}}}{\pi f_a}\frac{c_\gamma}{c_3}$.%

Our conservative bound extracted from Eq.~\eqref{eq:pigbound} by combining 8 TeV and 7 TeV LHC data together with Tevatron data sets the strongest existing limit on ALPs between 10 and 50 GeV: $f_a \gtrsim 500\text{ GeV}$, corresponding to $g_{a\gamma\gamma} \lesssim  10^{-5}$ GeV.
This is a major improvement with respect to the strongest existing bound in that range, which comes from measurements of $Z \to \gamma a$(jj) at LEP~I~\cite{Adriani:1992zm} giving $\text{BR}(Z\to \gamma +jj)<1-5\cdot 10^{-4}$. We checked that the other LEP limits in~\cite{Acciarri:1994gb,Abreu:393660,Abreu:1998yg} are not relevant for our choice of the anomalies. The limit from the boosted dijet search of CMS \cite{Sirunyan:2017nvi} is the strongest one between 50 and 65 GeV, while above 65 GeV the ATLAS~\cite{Aad:2014ioa} and CMS~\cite{CMS-PAS-HIG-17-013} diphoton searches take over.

The LHC has the potential to probe values of $f_a$ much larger than 1 TeV, as shown by the sensitivities lines in Fig.~\ref{fig:ALP}. The solid line  is obtained from Eq.~\eqref{eq:sensitivity} combining both 8 TeV and 7 TeV data with the finer possible binning. The dashed and dotted lines are the projected sensitivities respectively at LHC14 and HL-LHC, from 8 TeV and 7 TeV data, based on Eq.~\eqref{eq:sensitivityMC}.
Notice that the HL-LHC projection is stronger than the future ILC~\cite{Fujii:2017ekh} and FCC-ee~\cite{dEnterria:2016sca} reaches.
The latter is expected to probe $\text{BR}(Z\to \gamma +jj)\lesssim1-5\cdot 10^{-7}$, which correspond to $f_a\sim 1-3\text{ TeV}$ if $\mathcal{O}(10^{12})$ Z's will be produced.

The relative importance of low-mass diphoton bounds and sensitivities with respect to the other existing searches is robust with respect to choosing different values of the anomalies $c_{1,2,3}$, as long as $c_3 \neq 0$. For $c_{1,2} \gtrsim 4 c_3$, our conservative low-mass diphoton limit even overcomes the dijet exclusions between 50 and 65 GeV, while still doing largely better than LEP.

Other processes that could be relevant for an ALP with couplings as in Eq.~(\ref{eq:La123}) and mass above 10 GeV, like $Z\to 3\gamma$ at LEP (see e.g.~\cite{Mimasu:2014nea,Jaeckel:2015jla} for recent studies of this and other signatures), set limits that are too weak to even appear on the parameter space presented in Fig.~\ref{fig:ALP}. Analogously, the sensitivity of ALP searches in heavy ion collisions estimated in~\cite{Knapen:2016moh} is sizeably weaker than our conservative bounds.
The obvious reason is the generic suppression of the photon width compared to the gluon one by $(\alpha_{\rm em}/\alpha_s)^2$.
If Higgs decays to ALP pairs were allowed by the UV charge assignments, then the related constraints~\cite{Khachatryan:2016vau,Khachatryan:2017mnf,Aad:2015oqa} would apply. Their relative importance would be model dependent but in any case they would typically not probe $f_a$ values beyond a TeV, see~\cite{Bellazzini:2017neg} for more details.


As an exercise to conclude this section, we comment on the ALP interpretation of the excesses recently reported (both at $2.9\sigma$ local) by CMS in diphoton~\cite{CMS-PAS-HIG-17-013} and dijet~\cite{Sirunyan:2017nvi} searches, at invariant masses of 95 and 115 GeV respectively.
The ALP parameters that would fit each of them are
\begin{equation}
\frac{f_a}{c_\gamma} \simeq 470~{\rm GeV}\sqrt{ \frac{50~{\rm fb}}{\sigma_{\gamma\gamma}^{\text{sign}}} },\qquad c_3 \lesssim 2 \cdot c_\gamma\,,
\label{eq:excess_gaga}
\end{equation}
for the 95 GeV $\gamma\gamma$ excess, and
\begin{equation}
\frac{f_a}{c_3} \simeq 310~{\rm GeV}\sqrt{ \frac{300~{\rm pb}}{\sigma_{gg}^{\text{sign}}} },\qquad c_\gamma \lesssim 0.8\cdot c_3\,,
\label{eq:excess_jj}
\end{equation}
for the 115 GeV $jj$ one. 
$\sigma_{\gamma\gamma,gg}^{\text{sign}}$ are the theoretical 
signal cross sections of the excesses, whose normalization is chosen as follows. For the 95 GeV $\gamma\gamma$ excess we use the expected sensitivity at that mass as reported in Ref.~\cite{CMS-PAS-HIG-17-013}, for the 115 GeV $jj$ we use the analogous sensitivity reported in~\cite{Sirunyan:2017nvi} for a~$Z'$, and rescale it to an ALP produced in gluon fusion using Eq.~(\ref{eq:ggfromqq}).
Dijet bounds~\cite{Sirunyan:2017nvi} on the 95~GeV $\gamma\gamma$ excess~\cite{CMS-PAS-HIG-17-013}, and diphoton bounds~\cite{Aad:2014ioa} on the 115 GeV $jj$ excess~\cite{Sirunyan:2017nvi}, give the second inequalities in Eqs.~(\ref{eq:excess_gaga}) and (\ref{eq:excess_jj}) respectively.

Eqs.~(\ref{eq:excess_gaga}) and (\ref{eq:excess_jj}) allow to conclude that either of the two excesses, if coming from an ALP, could be interpreted in terms of reasonable values of $f_a$ and of the ABJ anomalies.
Such an ALP could be the first sign of a NP scale not too far from a TeV, still allowing the rest of the new states to be at $M_{\rm NP} \sim 4\pi f_a$ and hence out of the current LHC reach.

\section{Conclusions}


Theoretical frameworks such as Supersymmetry and Compositeness predict, on general grounds, the existence of pNGBs (ALPs) with couplings of relevance for colliders. Similar ALPs have also received much attention as mediators of Dark Matter interactions with the SM. The current experimental searches for these particles, however, still contain holes. In particular huge ($> 10^4$ pb) gluon fusion cross sections at the LHC, for ALP masses below 65 GeV, are allowed by all existing constraints.

In this paper, we used public data from inclusive diphoton cross section measurements at the LHC~\cite{Aad:2012tba,Aaboud:2017vol,Chatrchyan:2014fsa} to put a new bound on diphoton resonances between 10 and 65 GeV. We showed how this bound sets the by-far strongest existing constraint on the parameter space of ALPs that couple to both gluon and EW boson field strengths, see Fig.~\ref{fig:ALP}.
We have also derived indicative sensitivities that would be achievable by a proper LHC analysis, both with already existing 8 TeV data and at higher energies. 

We hope that this work will motivate the LHC collaborations to extend the mass range of their diphoton resonant searches to lower values.
Similar ideas could in principle be applied to probe light resonances decaying into other final states than diphotons. A great example is the current CMS search of boosted dijet resonances~\cite{Sirunyan:2017nvi}. Going to lower invariant masses in dijet -and perhaps in other- final states would certainly deserve further experimental effort.

\medskip

\subsection{Acknowledgements}

We thank Liron Barak, Sophia Borowka, Antonio Boveia, Marco Bonvini, Caterina Doglioni, Gabriele Ferretti, Mark Goodsell, Phil Harris, Bradley J. Kavanagh,  Greg Landsberg, Giovanni Marchiori, Pier Francesco Monni, Ian Moult, Kostantinos Vellidis and Andi Weiler for useful discussions. A special thank goes to Giovanni Marchiori, that first pointed us to~\cite{Aad:2012tba}. We also thank Antonio Boveia, Caterina Doglioni, Gabriele Ferretti, Ian Moult, David Shih, Lorenzo Ubaldi and Andreas Weiler for comments on the draft. D.R. thanks the LPTHE for kind hospitality during the completion of this work. F.S. is grateful to the Weizmann institute of Science, the Mainz Institute for Theoretical Physics (MITP), the Galileo Galilei Institute (GGI), and the Institut d'Astrophysique de Paris (IAP) for kind hospitality at various stages of this work.
\medskip

{\footnotesize
\noindent Funding and research infrastructure acknowledgements: 
\begin{itemize}
\item[$\ast$] A.M. is supported by the Strategic Research Program High Energy Physics and the Research Council of the Vrije Universiteit Brussel;
\item[$\ast$] F.S is partly supported by the European Research Council ({\sc Erc}) under the EU Seventh Framework Programme (FP7/2007-2013)/{\sc Erc} Starting Grant (agreement n.\ 278234 --- `{\sc NewDark}' project), by MITP, and by a {\sc Pier} Seed Project funding (Project ID PIF-2017-72);
\item[$\ast$] K.T. is supported in part by NSF award 1620628
\end{itemize}
}

\medskip

\appendix

\begin{figure*}[t!]
\includegraphics[width=0.44\textwidth]{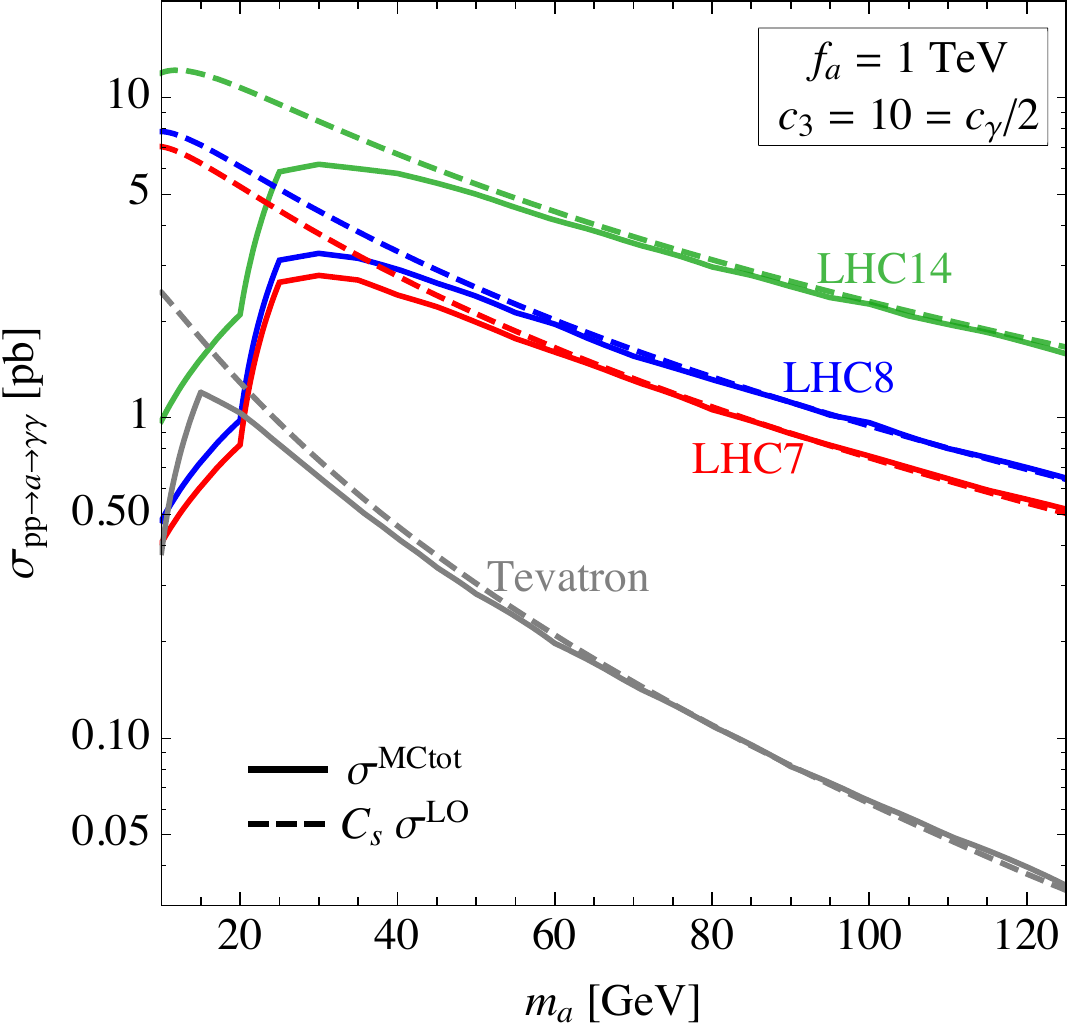}\qquad\qquad
\includegraphics[width=0.45\textwidth]{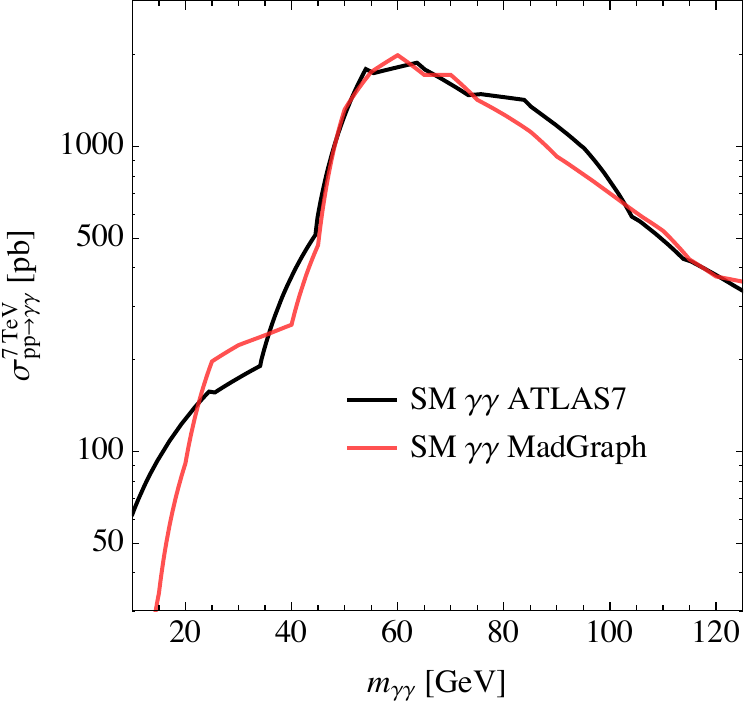}
\caption{Left: Total signal strengths from our MC simulation with minimal cuts (solid lines), compared with the LO theoretical signal strengths (dashed lines). See text for more details. Right: diphoton background shapes from our MC simulation (solid red) and from ATLAS cross section measurements (light blue) at 7 TeV.}
\label{fig:validation}
\end{figure*}

\section{Theoretical Signal Cross Sections \& Validation}\label{app:signalXS}
To compute the signal cross section we use
\begin{equation}
\sigma_{\gamma\gamma}^{\text{th}}(m_a, s)=\frac{K_\sigma}{K_{g}}\cdot \sigma_{\gamma\gamma}^{\text{LO}}(m_a, s)\,,
\end{equation}
where we work in the approximation $\Gamma_{\rm tot} \simeq \Gamma_{gg}$ (which is excellent in the parameter space that we have studied), and where
\begin{equation}
\sigma_{\gamma\gamma}^{\text{LO}}(m_a, s)=\frac{1}{m_a s} C_{gg}(m_a^2/s)\cdot\Gamma_{\gamma\gamma}\,, \label{eq:LO}
\end{equation}
%
%
\begin{equation}
C_{gg}=\frac{\pi^2}{8}\int^1_{m_a^2/s}\frac{dx}{x}f_g(x) f_g(\frac{m_a^2}{sx})\ ,\label{eq:PDFs}
\end{equation}
where $f_g(x)$ is the gluon PDF from the {\tt MSTW2008nnlo68} set~\cite{Martin:2009iq}, where we fix the pdf scale $q = m_a$.
We work with constant decay and production $K$-factors $K_{g} = 2.1$ and $K_{\sigma} = 3.7$. 
The former provides an approximation within $10\%$ of the most accurate expressions of~\cite{Djouadi:2005gj}, over the whole mass range of interest.
The latter was computed by using \emph{ggHiggs v3.5}~\cite{Ball:2013bra,Bonvini:2014jma,Bonvini:2016frm,Ahmed:2016otz} which includes full NNLO and approximate $\text{N}^{3}$LO corrections
, and where again we used the {\tt MSTW2008nnlo68} pdf set.
We find good agreement with the $K$-factor given in~\cite{Ahmed:2016otz}, for masses $m_a > 100$~GeV where their results are reported.
In principle $K_{\sigma}$ varies as a function of the center of mass energy and of the mass of the produced particle.  We find that the variation of $K_{\sigma}$ as a function of the mass for $40~{\rm GeV}<m_a<m_h$ is limited within $10\%$ of its central value, which is 3.7. Going from 8 TeV to 13 TeV does not lead to any sensible change in $K_{\sigma}$, while at 1.97 TeV $K_{\sigma}$ gets bigger by a factor of roughly $10\%$ which we neglect for simplicity.
Doing a more detailed estimate for masses below 40 GeV could require a full NLO simulation, which is beyond the scope of this paper. These approximations are more than sufficient for our purposes.  

Concerning the simulation of the signal and the derivation of the efficiencies, 
in Fig.~\ref{fig:validation} left we compare $\sigma^{\text{LO}}_{\gamma\gamma}$ with $\sigma^{\text{MCtot}}_{\gamma\gamma}$.
In the latter, in order to obtain the correct shape for the gluon fusion
cross section and to get the right $p_T$ distribution of
the extra jets, we considered matrix elements at parton
level with up to two extra jets in the final state and then
we matched them after parton shower to avoid double
counting \cite{Catani:2001cc,Krauss:2002up}.
$\sigma^{\text{LO}}_{\gamma\gamma}$ is the LO gluon fusion cross section from gluon PDF~\cite{Martin:2009iq} times LO diphoton branching ratio, the latter is the total diphoton signal strength obtained from MC simulation including only the minimal kinematical cuts on the two photons.
We see that, in the $m_a$ region where these cuts are not effective, $\sigma^{\text{LO}}_{\gamma\gamma}$ reproduces extremely well the $m_a$ shape of $\sigma^{\text{MCtot}}_{\gamma\gamma}$ upon rescaling it with a constant factor $C_s$. We find for the LHC $C_{7\,\text{TeV}}\simeq C_{8\,\text{TeV}}\simeq 0.85$ and for the Tevatron $C_{2\,\text{TeV}}\simeq 1$.

To have a validation of our procedure with a measured quantity, we simulate the SM diphoton background, matching it with the case of one and two extra jets. We then impose the kinematic and isolation cuts and verify that we are able to reproduce the shape \emph{and} size of the diphoton-only cross section measurements~\cite{Aad:2012tba,Aaboud:2017vol} of ATLAS, see Figure~\ref{fig:validation} right. The diphoton-only contribution is roughly $70\%$ of the total contribution, and the remaining $30\%$ is given by $\gamma\text{j}$ and jj final states (where the jet is faking a photon), that we do not include in our simulation nor in the experimental points with which we compare.





%
\begin{table*}[t!]
\centering
\begin{tabular}{|c|c|c|c|c|c|}
\hline
Experiment & Process &  Lumi &$\sqrt{s} $ & low mass reach& ref.   \\
LEPI& $e^{+}e^{-}\to Z\to \gamma a\to\gamma jj $&12 pb$^{-1}$& Z-pole&$10\text{ GeV}$&\cite{Adriani:1992zm}\\
LEPI& $e^{+}e^{-}\to Z\to \gamma a\to\gamma \gamma\gamma $&78 pb$^{-1}$ & Z-pole&$3\text{ GeV}$&\cite{Acciarri:1994gb}\\
LEPII& $e^{+}e^{-}\to Z^{*},\gamma^{*}\to \gamma a\to \gamma jj $&9.7,10.1,47.7 pb$^{-1}$& 161,172,183\text{ GeV}&$60\text{ GeV}$&\cite{Abreu:393660}\\
LEPII& $e^{+}e^{-}\to Z^{*},\gamma^{*}\to \gamma a\to \gamma \gamma\gamma $&9.7,10.1,47.7 pb$^{-1}$& 161,172,183\text{ GeV}&$60\text{ GeV}$&\cite{Abreu:393660,Abreu:1998yg}\\
LEPII& $e^{+}e^{-}\to Z^{*},\gamma^{*}\to Z a\to jj\gamma\gamma $&9.7,10.1,47.7 pb$^{-1}$& 161,172,183\text{ GeV}&$60\text{ GeV}$&\cite{Abreu:393660}\\
D0/CDF& $p\bar{p}\to a\to \gamma\gamma$&7/8.2 fb$^{-1}$&1.96\text{ TeV}&$100\text{ GeV}$ &\cite{Bland:2011dc}\\
ATLAS & $p p\to a\to \gamma\gamma$& 20.3 fb$^{-1}$&8\text{ TeV}& $65\text{ GeV}$&\cite{Aad:2014ioa}\\
CMS &$p p\to a\to \gamma\gamma$&19.7 fb$^{-1}$&8\text{ TeV}& $80\text{ GeV}$&\cite{CMS-PAS-HIG-14-037}\\
CMS  &$p p\to a\to \gamma\gamma$&19.7 fb$^{-1}$&8\text{ TeV}& $150\text{ GeV}$&\cite{Khachatryan:2015qba}\\
CMS  &$p p\to a\to \gamma\gamma$& 35.9 fb$^{-1}$&13\text{ TeV}& $70\text{ GeV}$&\cite{CMS-PAS-HIG-17-013}\\
\hline
\hline
CMS& $pp\to a\to jj$& $18.8\text{ pb}^{-1}$& $8\text{ TeV}$& $500\text{ GeV}$&\cite{Khachatryan:2016ecr}\\
ATLAS& $pp\to a\to jj$&20.3 fb$^{-1}$&8\text{ TeV}&$350\text{ GeV}$&\cite{Aad:2014aqa}\\
CMS& $pp\to a\to jj$& $12.9\text{ pb}^{-1}$& $13\text{ TeV}$& $600\text{ GeV}$&\cite{Sirunyan:2016iap}\\
ATLAS& $pp\to a\to jj$& 3.4 fb$^{-1}$&13\text{ TeV}&$ 450\text{ GeV}$&\cite{ATLAS-CONF-2016-030}\\
CMS& $pp\to j a\to jjj$& $35.9\text{ pb}^{-1}$& $13\text{ TeV}$& $50\text{ GeV}$& \cite{Sirunyan:2017nvi}\\
\hline
\hline
UA2& $p\bar{p}\to a\to \gamma\gamma$& $13.2\text{ pb}^{-1}$& $0.63\text{ TeV}$& $17.9\text{ GeV}$& \cite{Alitti:1992hn}\\
D0& $p\bar{p}\to a\to \gamma\gamma$&4.2 fb$^{-1}$& 1.96\text{ TeV}&$8.2\text{ GeV}$&\cite{Abazov:2010ah}\\
CDF & $p\bar{p}\to a\to \gamma\gamma$&5.36 fb$^{-1}$& 1.96\text{ TeV}&$6.4\text{ GeV}$&\cite{Aaltonen:2011vk,PhysRevLett.110.101801}\\
ATLAS & $p p\to a\to \gamma\gamma$&4.9 fb$^{-1} $& 7\text{ TeV}&$9.4\text{ GeV}$&\cite{Aad:2012tba}\\
CMS  &$p p\to a\to \gamma\gamma$&5.0 fb$^{-1}$ & 7\text{ TeV} &$14.2\text{ GeV}$&\cite{Chatrchyan:2014fsa}\\
ATLAS & $p p\to a\to \gamma\gamma$&20.2 fb$^{-1} $& 8\text{ TeV}& $13.9\text{ GeV}$&\cite{Aaboud:2017vol}\\
\hline
\end{tabular}
\caption
{In the top of the Table we list the relevant searches involving at least a photon in the final state at different colliders, and lowest value of invariant mass that they reach. In the middle we also include the most recent LHC dijet searches (see Ref.~\cite{Dobrescu:2013coa} for a list of older searches). On the lower part of the Table we summarize the available diphoton cross section measurements with their minimal invariant mass reach, which we estimate from the minimal $p_T$ cuts on the leading and subleading photon and the isolation cuts of the diphoton pair. \label{tab:summary_searches}}
\end{table*}
%

\begin{figure*}[t!]
\includegraphics[width=0.45\textwidth]{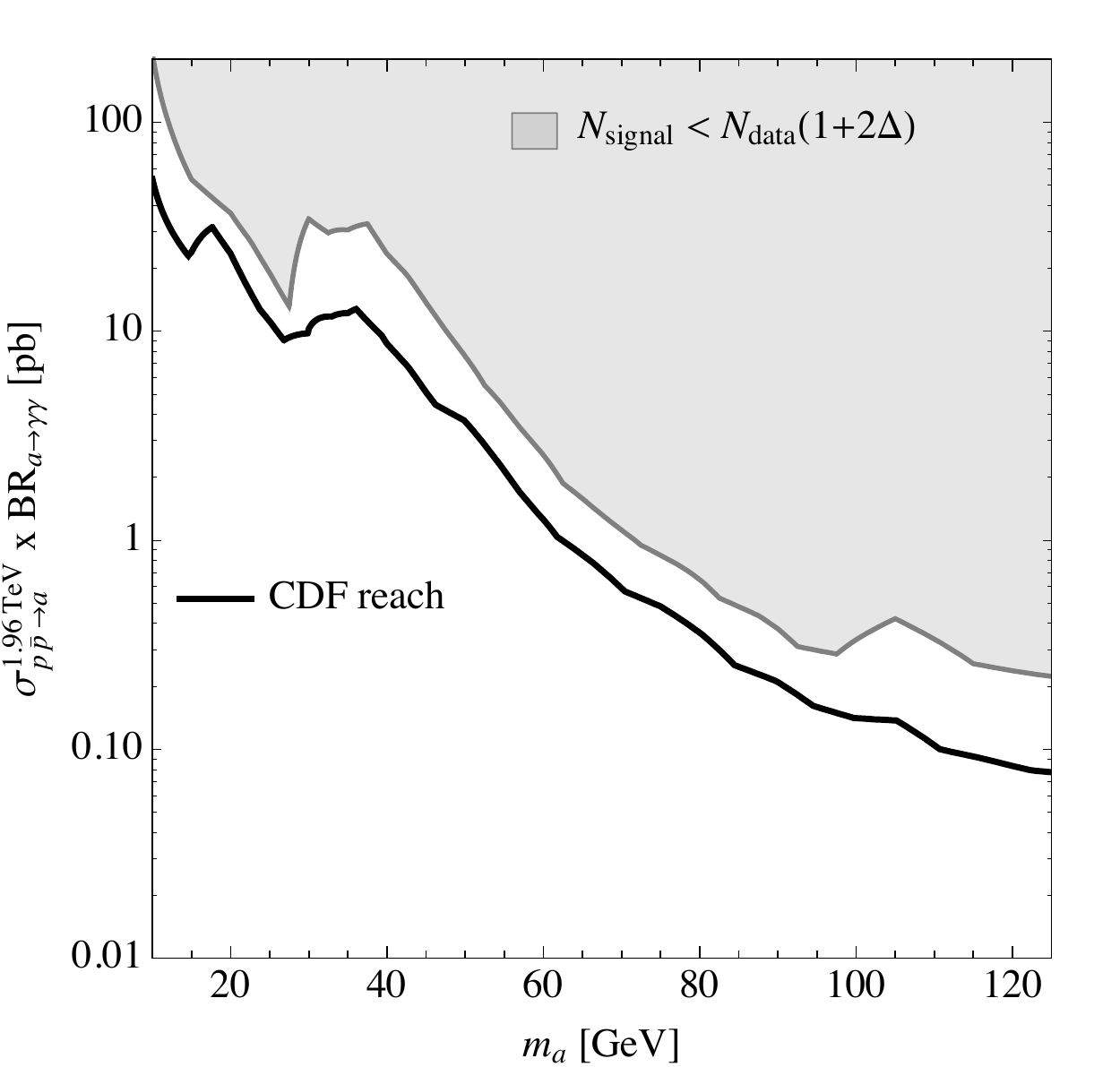}\qquad
\includegraphics[width=0.48\textwidth]{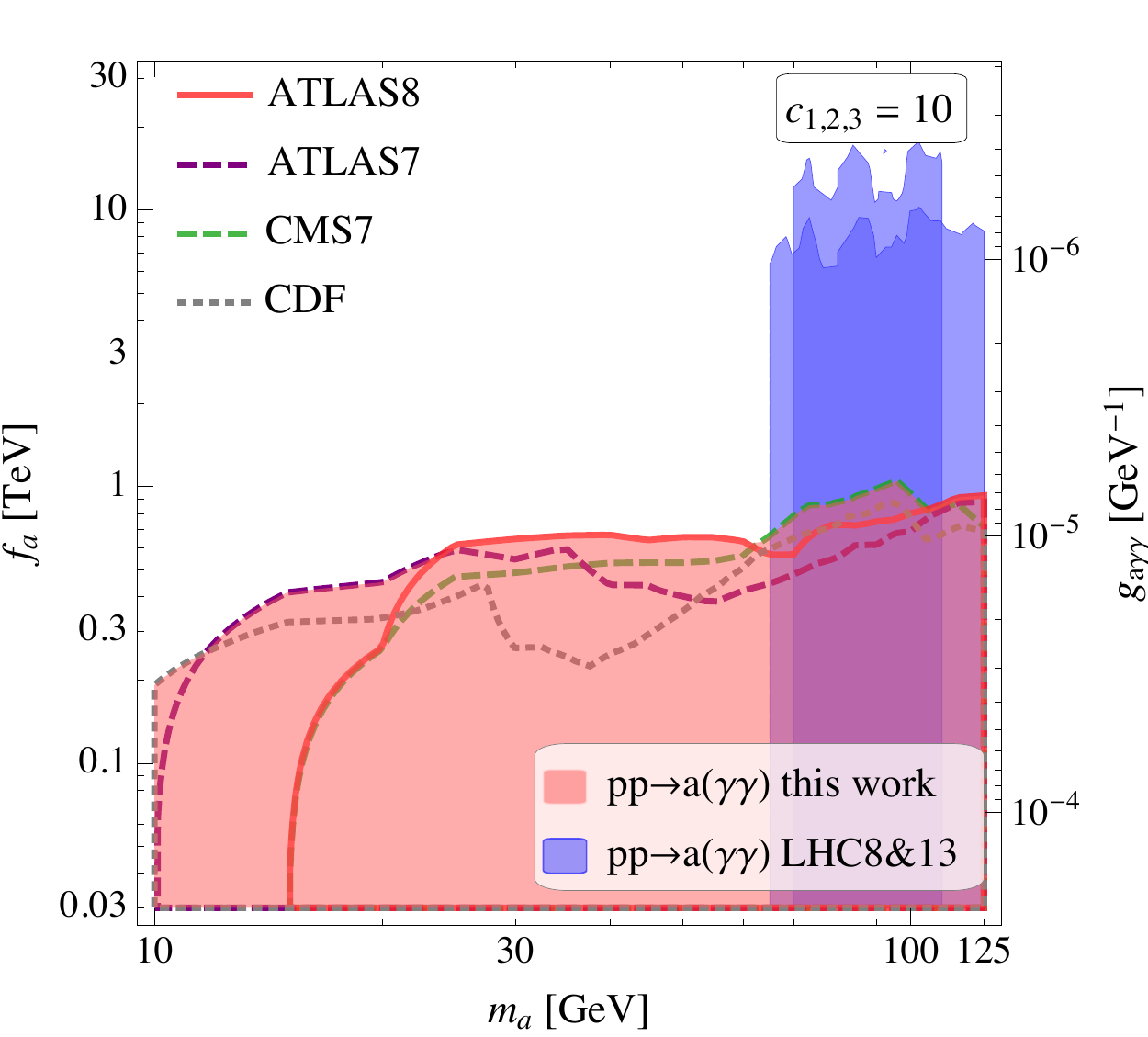}
\caption{Left: Bound (shaded) and expected sensitivity after rebinning (lines) on the diphoton signal strength of a resonance produced in gluon fusion at the Tevatron. Right: Unfolding of the bound in the ALP parameter space extracted from the different diphoton cross section measurements. The final bound (pink shaded region) is the union of the ATLAS data at 8 TeV \cite{Aaboud:2017vol} (pink solid) and at 7 TeV \cite{Aad:2012tba} (purple dashed), of the CMS data at 7 TeV \cite{Chatrchyan:2014fsa} (green dashed) and of the CDF data at 1.96 TeV \cite{Aaltonen:2011vk,PhysRevLett.110.101801} (grey dotted).}
\label{fig:extra1}
\end{figure*}

\section{Rebinning}
\label{app:binning}
We specify here the procedure we follow to reduce the bin size down to the invariant mass resolution of the ECAL for every experiment. The CDF and D0 energy resolutions are 
\begin{align*}
&\text{ CDF:  }&\frac{\delta E_\gamma}{E_\gamma}=13.5\%\cdot \left(\frac{\text{GeV}}{E_\gamma}\right)^{1/2}\, ,\\
&\text{ D0:  }& \frac{\delta E_\gamma}{E_\gamma}=18\%\cdot \left(\frac{\text{GeV}}{E_\gamma}\right)^{1/2}\ .
\end{align*}
The CDF energy resolution is derived from~\cite{Balka:1987ty,Hahn:1987tx}. Using the same formula we can extrapolate the resolution of D0 at different energies given that in \cite{Aaboud:2017vol} they quote a resolution of $3.6\%$ for $E_\gamma=50\text{ GeV}$. 
The ATLAS and CMS ECAL energy resolutions are extracted from~\cite{2006NIMPA.568..601A} and \cite{deFavereau:2013fsa}, and read
\begin{align*}
&\text{ ATLAS:  }&\frac{\delta E_\gamma}{E_\gamma}&=10\% \left(\frac{\text{GeV}}{E_\gamma}\right)^{1/2}\oplus0.7\%\, ,\\
&\text{ CMS:  }& \frac{\delta E_\gamma}{E_\gamma}&=7\% \left(\frac{\text{GeV}}{E_\gamma}\right)^{1/2}\oplus 35\%\frac{\text{GeV}}{E_{\gamma}}\oplus0.7\%\, .
\end{align*}
These are related to  smearing of the diphoton resonance. The invariant mass can be written as $m_{\gamma\gamma}^2=2 E_{\gamma_1} E_{\gamma_2}(1-\cos\Delta\theta)$ where $\Delta\theta$ is the angle between the 2 photon momenta. 
An appropriate bin size that contains 95\% of the signal is obtained by an interval of $m_{\gamma\gamma}\pm 2\delta m_{\gamma\gamma}$ where 
\begin{equation}
\frac{\delta m_{\gamma\gamma}}{m_{\gamma\gamma}}\approx\frac{1}{2}\left(\frac{\delta E_{\gamma_1}}{E_{\gamma_1}}\oplus \frac{\delta E_{\gamma_2}}{E_{\gamma_2}}\right)\ .\label{smearing}
\end{equation}
For $m_a>E_{\gamma_1}+E_{\gamma_2}$ we can neglect any possible boost coming from extra radiation. 
Then, as a cross-check of Eq.~\eqref{smearing}, we apply it to the $125\text{ GeV}$ Higgs with $E_{\gamma_1}=E_{\gamma_2}=m_{\gamma\gamma}/2$ and get the Gaussian smearing of $\delta m_{\gamma\gamma} =1.27\ (1.11)\text{ GeV}$ for ATLAS (CMS) which is in the same ballpark of the detector smearing effects reported in the ATLAS  \cite{Aad:2012tfa} and CMS \cite{Chatrchyan:2012xdj} analysis. 
Also the mass dependence of the smearing provided by ATLAS in~\cite{Aad:2014ioa} is reproduced by Eq.~\eqref{smearing}. For $m_a<E_{\gamma_1}+E_{\gamma_2}$, the trigger threshold on the two photons energies sets the lower limit on the bin size which is $\approx 2.9\text{ GeV}$ for the 7 TeV ATLAS analysis, $\approx 23.6 \text{ GeV}$ for the 8 TeV ATLAS analysis and $\approx 3.3 \text{ GeV}$ for the 7 TeV CMS analysis.

\section{More Details on Cross Section Measurements}\label{app:previoussearches}
In Table~\ref{tab:summary_searches} we summarize for completeness the existing collider analysis targeting final states with at least one photon.
We also include the most recent dijet resonance searches at the LHC, while we refer to Ref.~\cite{Dobrescu:2013coa} for a collection of previous searches involving purely hadronic final states.
 
In the following we then report the detailed cuts of the cross section measurements at the Sp$\bar{\rm p}$S, Tevatron and the LHC.
\begin{itemize}
\item[$\circ$] In the UA2 analysis \cite{Alitti:1992hn} diphotons events are required to have $p_{T_1}>10\text{ GeV}$ and $p_{T_2}>9\text{ GeV}$. The extra cut on $Z\equiv -\frac{p_{T_1}\cdot p_{T_2}}{\vert p_{T_1}\vert^2}>0.7$ selects photon pairs almost back to back ($\cos\Delta\phi\lesssim 0.78$). As a consequence, given that $m_{\gamma\gamma}^2= 2 p_{T1} p_{T2} (\cosh\Delta\eta - \cos\Delta\phi)$, we find that the invariant mass reach can only go down to $m_{\gamma\gamma}\gtrsim 17.9\text{ GeV}$. 
\item[$\circ$] In the CDF analysis \cite{Aaltonen:2011vk,PhysRevLett.110.101801} two isolated photons with $p_{T_1}>15\text{ GeV}$ and $p_{T_2}>17\text{ GeV}$ respectively are required to be reconstructed within the geometrical acceptance of the electromagnetic calorimeter (ECAL) $0.05<\vert\eta\vert<1.05$ with angular separation $\Delta R\equiv\sqrt{\Delta\eta^2+\Delta\phi^2}$ greater than $0.4$. The binning of the data in the diphoton invariant mass is constant and equal to $5\text{ GeV}$ in the mass range of interest. The bin with $10\text{ GeV}<m_{\gamma\gamma}<15\text{ GeV}$ of the CDF analysis has an anomalously low inclusive cross section of 0.004 pb which is one order of magnitude smaller than the ones in the adjacent bins. This feature is not present in the cross section measurements for photons with $p_T>m_{\gamma\gamma}$. Since our signal will be not affected by this extra $p_T$ cut we decided to conservatively include this latter experimental point in our bound. A more careful understanding of the Tevatron data would be required to be confident that the lower mass bins are not affected by extra large systematics.\footnote{We thank Konstantinos Vellidis for correspondence on the features of the CDF data and for pointing us to the most updated reference \cite{PhysRevLett.110.101801}.} 

\item[$\circ$] In the D0 analysis \cite{Abazov:2010ah} two isolated photons are required to have $p_{T_1}>21\text{ GeV}$ and $p_{T_2}>20\text{ GeV}$ respectively and to be within  $\vert\eta\vert<0.9$ with angular separation $\Delta R>0.4$. The binning of the data in the diphoton invariant mass is $15\text{ GeV}$ below 50 GeV and 10 GeV above. 

\item[$\circ$] The $7$ TeV ATLAS analysis \cite{Aad:2012tba} requires two isolated photons with $p_{T_1}>25\text{ GeV}$ and $p_{T_2}>22\text{ GeV}$ respectively with $\Delta R>0.4$ and within the geometrical acceptance of the ECAL ($\vert\eta\vert<1.37$ and $1.52<\vert\eta\vert<2.37$). Tight isolation and selection criteria on the photons are imposed using  the standard DELPHES ATLAS card.\footnote{
We checked that changing the isolation cuts from the standard DELPHES ATLAS to the curve given in \cite{Aad:2010sp} does not modify the efficiency by more than $\sim 10\%$. The isolation cuts for the Tevatron are instead approximately imposed by using the standard DELPHES CMS card.
}
The first bin takes $0<m_{\gamma\gamma}<20 \text{ GeV}$ while the width of all the further bins is 10 GeV.

\item[$\circ$] In the $7$ TeV CMS analysis \cite{Chatrchyan:2014fsa} two isolated photons with $p_{T_1}>40\text{ GeV}$ and $p_{T_2}>25\text{ GeV}$ are required to be reconstructed in the pseudorapidity range $\vert\eta\vert<2.7$ with $\vert\eta\vert\notin[1.44,1.57]$ and with angular separation $\Delta R>0.45$. The photon isolation is imposed by using the standard DELPHES CMS card. The first very wide bin takes $0<m_{\gamma\gamma}<40 \text{ GeV}$ while the other up to 120 GeV have a variable width between 10 and 20 GeV.

\item[$\circ$] The $8$ TeV ATLAS analysis \cite{Aaboud:2017vol} requires two isolated photons with $p_{T_1}>40\text{ GeV}$ and $p_{T_2}>30\text{ GeV}$ with angular separation $\Delta R>0.4$. The geometrical acceptance of the electromagnetic calorimeter is extended to $\vert\eta\vert<1.37$ and $1.56<\vert\eta\vert<2.37$ and tight isolation and selection criteria are also imposed following the standard DELPHES ATLAS card. The first bin has $0<m_{\gamma\gamma}<30 \text{ GeV}$ then there are two bins with a 20 GeV width up to $m_{\gamma\gamma}=70\text{ GeV}$ and all the other bins have a width of 10 GeV. 
\end{itemize}

The resulting efficiencies for the signal at the different experiments are reported in Table~I of the main letter. For completeness we report the model independent bound obtained from CDF cross section measurement in Fig.~\ref{fig:extra1} left. Notice that the effect of the rebinning is marginal in this case because of the already fine binning of the experimental data. We checked that the bound extracted from D0 data is always weaker than the CDF one and we do not plot it for simplicity.

We are now ready to compare the bounds obtained from the different cross section measurement in Fig.~\ref{fig:extra1} right. As we see the cross section measurements at the LHC lead to a stronger bound than CDF besides for very low masses $m_a\sim 10\text{ GeV}$. Indeed the very low $p_T$ cuts of the CDF analysis allow to have a $\sim1\permil$ efficiency for $m_a=10\text{ GeV}$ (Table~I of the main letter), as opposed to the zero efficiency of the other experiments. Even if the CDF data would have larger systematics in the lower bins, as suggested by the strange feature discussed in the bullet points above, our bound would not be modified by much.

\begin{figure*}[t!]
\includegraphics[width=0.49\textwidth]{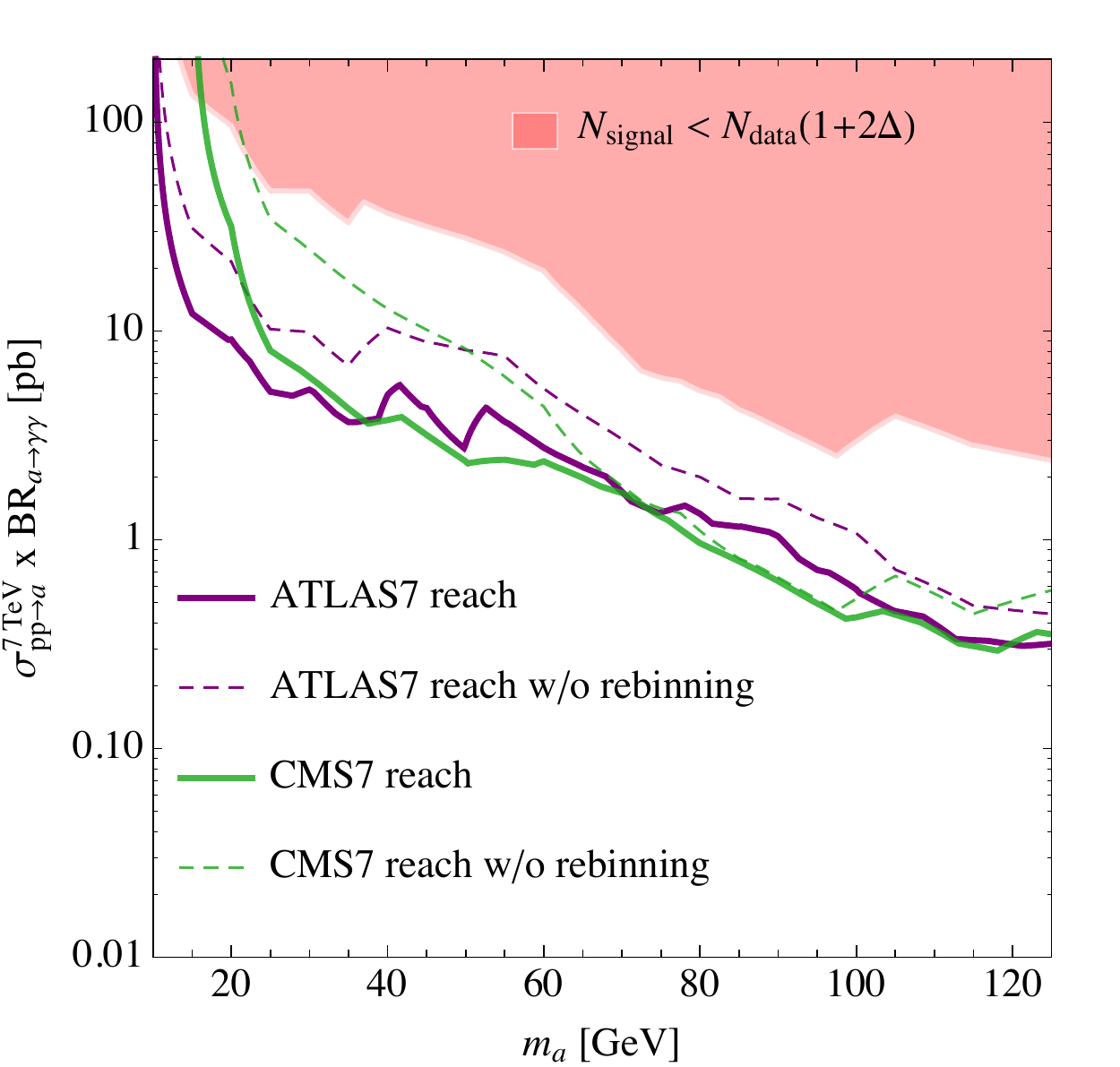}
\includegraphics[width=0.49\textwidth]{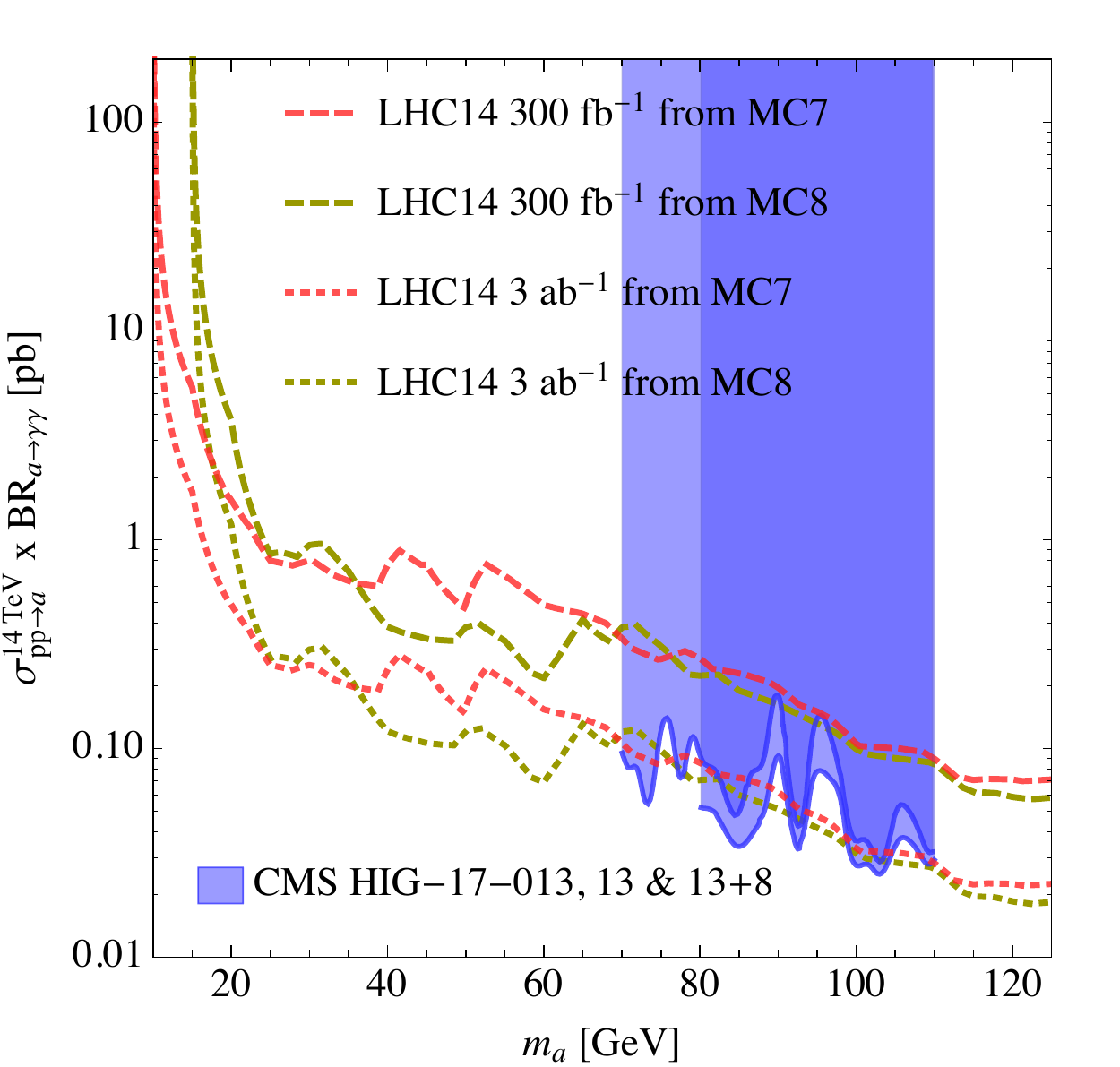}
\caption{
Bounds (shaded) and expected sensitivities (lines) on the diphoton signal strength of a resonance produced in gluon fusion, at 7 (left) and 14 (right) TeV. The bounds at 14 TeV are rescaled from the lower energy CMS ones. See text for more details.}
\label{fig:modelind2}

\end{figure*}

\section{7 TeV data \& projections at 14~TeV}\label{app:extramaterial}

For completeness we present here our results based on ATLAS and CMS 7 TeV data \cite{Chatrchyan:2014fsa,Aad:2012tba} and our projections at LHC14 and HL-LHC.

The conservative bound at 7 TeV derived is shown in the left panel of Fig.~\ref{fig:modelind2}, and it extends to a lower invariant mass with respect to one based on 8 TeV data. This can be explained by noticing that the ATLAS measurement at 7 TeV has the lowest invariant mass reach given the low $p_T$ cuts on the two photons (see Table~\ref{tab:summary_searches}).
In Fig.~\ref{fig:modelind2} left we also show the 7 TeV projections based on Eq.~(6) of the main letter, with the original binning of the experimental measurements (dashed) and with the finest possible binning allowed by ECAL resolution (solid).  

In Fig.~\ref{fig:modelind2} right we show the projections at LHC14 (dashed) and HL-LHC (dotted) based on Eq.~(7) of the main letter, and taking as initial input $\sigma^{\text{sens}}_{\gamma\gamma, \rm low}$ the ATLAS7 and ATLAS8 sensitivities determined from Eq.~(6) of the main letter.
The agreement far from the cuts between the sensitivities projected from 7 and 8 TeV measurements %
is a nice consistency check of our procedure. The comparison between our projections and the present bounds from the recent CMS search at 13 TeV \cite{CMS-PAS-HIG-17-013} (which we have rescaled to 14 TeV as the Higgs boson production cross section) shows that an actual search at for low-mass diphoton resonances could certainly do better than our crude estimates.

\bibliography{LightRes}
\end{document}